\documentclass[11pt,reprint]{iopart}
\usepackage{amsbsy}
\usepackage{amssymb}   
\usepackage{graphicx}
\usepackage{bm}
\usepackage{lineno}

\def\dd{\mbox{d}}

\def\R{{\bf R}}

\def\r{{\bf r}}

\begin{document}

%Submitted to Journal of Physics A (2006)

\title[Chemotactic Target Selection]{Frequency-dependent Chemolocation and Chemotactic Target Selection}

\author{Sarah A Nowak$^{1}$, B Chakrabarti$^{1}$, Tom
Chou$^{1,2}$ and Ajay Gopinathan$^{3}$}
\address{$^{1}$Dept.~of Biomathematics, UCLA,
Los Angeles, CA 90095-1766}
\address{$^{2}$Dept.~of Mathematics, UCLA,
Los Angeles, CA 90095-1555}
\address{$^{3}$ School of Natural Sciences, University of California, Merced, CA 95344}
\ead{tomchou@ucla.edu}

\date{\today}

%\runninglinenumbers*

\begin{abstract}
Chemotaxis is typically modeled in the
context of cellular motion towards a static, exogenous source of
chemoattractant. Here, we propose a time-dependent mechanism of
chemotaxis in which a self-propelled particle ({\it e.g.}, a cell)
releases a chemical that diffuses to fixed particles (targets) and
signals the production of a second chemical by these targets.  The
particle then moves up concentration gradients of this second
chemical, analogous to diffusive echolocation.
% The mathematical problem is instinsically nonlinear due to a moving
% source. We use both dirent numerical integration of the governing equations as well
% as evaluation of Green's function to find cell trajectories. 
% The chemotatctic behavior of cells is numerically explored as a function of
% parameters such as chemoattractant reponse and probe chemical release frequency.
%
%
%
When one target is present, we describe probe release strategies that
optimize travel of the cell to the target. In the presence of multiple
targets, the one selected by the cell depends on the strength and,
interestingly, on the frequency of probe chemical release.  Although
involving an additional chemical signaling step, our chemical
``pinging'' hypothesis allows for greater flexibility in regulating
target selection, as seen in a number of physical or biological
realizations.
\end{abstract}

\maketitle

\section{Introduction}

\noindent Organisms that employ chemical signaling for functions such
as antimicrobial defense mechanisms~\cite{Chet:71,Alberts:02,Cain:03}
and nutrient uptake~\cite{Bray:92,BERG} must coordinate the influences
of a complicated spatio-temporal mixture of signaling molecules
emanating from possibly many sources.  A fundamental problem in
chemotaxis is how the cell determines a strategy to best select a
specific target from many?  This problem arises for many chemotactic
organisms, such as {\it E. coli}~\cite{Mittal:03}, {\it Myxococcus
xanthus} ~\cite{Velicer:03}, and {\it Dictyostelium
discoideum}~\cite{Tyson:89}, and poses a challenging theoretical
question.

%A theoretical model of such targeting strategies is important in
%designing robotic communications~\cite{Ogras:04}, probing for natural
%resources, understanding antimicrobial defense
%mechanisms~\cite{Chet:71,Alberts:02,Cain:03} and signaling among
%cells~\cite{Bray:92}.  

%Although targets may be physically discerned
%through acoustical, optical, thermal, or chemical signals, at the
%cellular level, directed movement usually proceeds via the sensing of
%local chemical gradients~\cite{Adler:66,Berg:All}.  Chemotaxis is key
%for many organisims, including bacteria such as {\it Myxococcus
%xanthus}~\cite{Velicer:03}, {\it Escherichia coli}~\cite{Mittal:03},
%and {\it Dictyostelium discoideum}~\cite{Tyson:89}.

Here, we propose a biologically plausible mechanism involving active
sensing, in which chemical signaling is initiated by a chemical prompt from the
chemotactic cell. Potential biological manifestations of active
sensing include ``diffusion sensing'' \cite{DS} and cancer cell
chemotaxis~\cite{MMP1}. Diffusion sensing is a particular variant of
quorum sensing that involves release of a metabolically inexpensive
compound, inducing nearby cells to emit the main signal back to the
first cell indicating to it that that other chemically responsive
targets are in the vicinity~\cite{DS}.  Cancer cells also exploit this
type of indirect sensing by producing excess amounts of matrix
metalloproteinases (MMP) which cleave substrates (such as growth
factors and laminins) bound to the extracellular matrix
(ECM)~\cite{MMP1}.  These cleavage products diffuse back to the cancer
cells, inducing their chemotaxis. Dynamic multistep chemotaxis
mechanisms may also arise in the mating of yeast, where each sex of
yeast emits its own pheromone that up-regulates gene expression of the
complementary pheromone in the opposite sex~\cite{YEAST}.

In our model of active sensing, moving chemical sources and targets
interact through the time-dependent diffusion of signaling molecules
Therefore, the timing of the release and detection of chemoattractants
will also be important in the overall chemotactic process
\cite{DEGENNES}. Indeed, there is evidence that the probe emission of
MMP's is transcriptionally regulated and is
time-dependent~\cite{MMP2}. Experiments have also demonstrated the
existence of robust and tunable oscillations in transcription in {\it
E. coli} \cite{STRICKER} and mammalian cells \cite{TIGGES}, and cAMP
release in {\it Dictyostelium discoideum} \cite{Tyson:89,DICTY}. Another
important virtue of the proposed dynamic sensing mechanism is that it
allows cells to detect local, transient chemoattractant gradients in
complex media where steady-state gradients cannot be sustained. For
example, branched, ``dead-end'' volumes with impenetrable boundaries
cannot support a steady-state chemical gradient, but can allow a
transient gradient. Therefore, chemotaxis under time-dependent, and in
particular, time periodic conditions are realizable systems for
further exploration.

%These MMP's that are released then can interact and cleave and degrade
%several extracellular matrix (ECM) substrates. These include things
%like growth factors and laminin among others that are known to induce
%chemotaxis in these cancer cells.
%Thus the MMP's are like chemical A released by the cell and then
%interaction with the ECM (target) releasing chemoattractants like
%growth factors (B) that the cell chemotaxxes in response to. In short
%almost exactly our system. Is the MMP production time dependent?
%Dont know but it is transcriptionally regulated so it is possible
%there is some oscillatory behavior. In any case we can say that our
%model is motivated by this system leading us to the general question
%of cells that create their own gradients via interaction with the
%environment and we further consider interesting time dependent
%phenomena.  Finally the last one is some lame model to show
%oscillations in MMP expression though the time scales are in hours and
%so this is probably not useful but it is useful as a reference.

\begin{figure}[t]
\begin{center}
\includegraphics[width=3.5in]{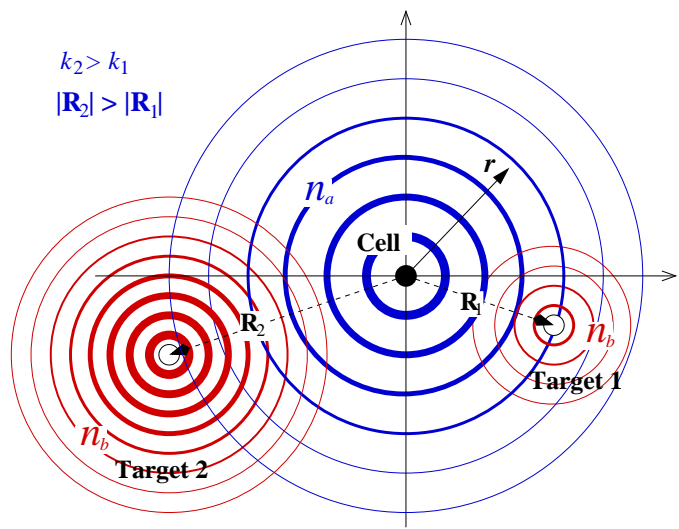}
\end{center}
\caption{\baselineskip=12pt A schematic of a motile cell releasing probe chemical $a$
(blue) which gets converted at fixed targets to a chemoattractant $b$
(red) which the cell detects and uses to move. For the 
two target case shown here, the initially farther Target 2,
has a larger $a\rightarrow b$  conversion rate ($k_2 >  k_1$).}
\label{Schematic-Fig}
\end{figure}

Basic models for these potentially novel time-periodic chemotactic
systems are currently lacking.  Here, we distill the extracellular
components of a dynamic multistep chemotaxis mechanism into an
essential physical model that describes gradient-guided cellular
motion towards chemically responsive targets.  Classic chemotaxis
models by Keller and Segel~\cite{Keller:70}, and their
extensions~\cite{DEGENNES,Grima:05} consider passive sources of a
chemoattractant to which a cell responds, or consumption of a passive
chemical \cite{DIL95}.  Contrary to the classic model of chemotaxis
where the cell moves along a static chemical gradient of nutrients
already present in the environment, we consider the following
scenario. Initially, a cell sends out a chemical probe signal (species
$a$), which diffuses to stationary targets. Upon contact with the
targets, the probe chemical $a$ induces the target to release a
different chemical (species $b$), which diffuses back to the cell. In
response, the cell moves up the chemical gradient of the
chemoattractant $b$.  A schematic of our proposed ``chemolocation''
mechanism, a diffusive analogue of sonar, is shown in
Fig.~\ref{Schematic-Fig}, where one cell and two targets are
depicted. Here, we assume that chemoattractant $b$ is different from
probe chemical $a$ (``paracrine'' signaling) and that the cell uses
only $b$ to guide its motion towards the target(s).

%, although chemical are known to serve as both
%parachrine and autochrine signals.
%~\cite{parachrine}.

%The ``paracrine''
%aspect of our model arises if production of the complementary
%pheromone is increased. An ``autocrine''~\cite{autochrine} mechanism
%(where $b=a$) is relevant to bacterial swarming and formation of
%fruiting bodies~\cite{Bonner:83,Gerisch:87}, and can also be described
%within our mathematical framework.
%%%%%%%%%%%%%%%%%%%%%%%%%%%%%%%
%%%%%%%%%%%%%%%%%%%%%%%%%%%%%%%

\section{Mathematical Model}
Denote the concentrations at spatial position $\r$ of probe chemical
$a$ emitted by the cell, and chemoattractant $b$ produced by the fixed
targets by $n_{a}(\r,t)$ and $n_{b}(\r,t)$, respectively. The
production of chemoattractant $b$ by the targets may be initiated by
binding of probe $a$ to receptors on the targets.  Although the cell
needs to detect spatial gradients in $n_{b}$, we assume for simplicity
that both the cell and the target can be treated as point particles
when considering the diffusive dynamics of $n_{a}$ and $n_{b}$.

The governing equations in our model are
\begin{equation}\label{DIFFA}
\dot{n}_{a}(\r,t)= D_{a}\nabla^{2} n_{a} - \mu_{a} n_{a} + F(t)
\delta(\r-\R(t)),
\end{equation}
\begin{equation}\label{DIFFB}
\dot{n}_{b}(\r,t) = D_{b}\nabla^{2} n_{b} - \mu_{b} n_{b} + 
\sum_{j}\!\delta(\r - \R_{j})K_{j}\!\left[n_{a}(\r,t-t'),t\right],
\end{equation}
\begin{equation}
\dot{\bf R}(t) = \int_{-\infty}^{t}\gamma(t')\nabla U[n_{b}(\r,t-t')]\dd t'\big|_{\r = \R(t)}
%
%\dot{\R}(t) = %\g\!\!\! \lim_{\r\rightarrow\R(t)}\!\! 
%\gamma \nabla U\left[n_{b}(\r,t)\right]\big|_{\r = \R(t)}, 
\label{RDOT}
\end{equation}

\noindent where $\delta(\cdot)$ is the Dirac delta function, $\R(t)$
is the position of the moving cell, $\R_{j}$ are the fixed target
positions, $D_{a}, D_{b}$ are the uniform probe and chemoattractant
diffusivities, and $\mu_{a}, \mu_{b}$ their uniform degradation
rates. In Eq.~\ref{DIFFA}, $F(t)$ represents the time-dependent
emission of probe chemical $a$ by the cell at position $\R(t)$, while
in Eq.~\ref{DIFFB} the term
$\sum_{j}\delta(\r-\R_{j})K_{j}\left[n_{a}(\r,t-t'),t\right]$
represents the total production of $b$ from $a$ by all the targets
$j$, each at fixed position $\R_{j}$.  Note that for systems in
dimension $d\leq 2$, we must assume $\mu_{a},\mu_{b}>0$ if
$F(t\rightarrow \infty) > 0$ in order for the chemical concentrations
to remain bounded. The functional
$K_{j}\left[n_{a}(\R_{j},t-t'),t\right]$ embodies all chemical steps
in the production of chemoattractant $b$ in response to contact with
probe chemical $a$. The production of chemoattractant $b$ occurs with
a delay $t'$, (the time taken for a single or multi-step reaction
along the signaling pathway) after the targets are exposed to the
probe $a$ at a time $t -t'$ ~\cite{Bray:92,TU}. We will henceforth
assume the simplest model for the kernel $K_{j}\left[n_{a}(\R_{j},
t-t'), t \right] = k_{j} n_{a}(\R_{j}, t)$, representing instantaneous
production, with rate $k_{j}$, of chemoattractant proportional to the
concentration of probe.

Equation \ref{RDOT} describes the motion of the cell in an effective
time-dependent potential $U\left[n_{b}(\R(t), t) \right]$ generated by
the dynamics of the chemoattractant $n_{b}(\R(t),t)$. The functional
$U\left[n_{b}(\r, t)\right]$ could be a nonlinear function of
$n_{b}(\r,t)$, such as one with threshold and saturation like $U
\propto n_{b}^{\alpha}(\r,t)/(\textrm{const.} +
n_{b}^{\alpha}(\r,t))$, where $\alpha$ is the Hill coefficient. This
form may be more appropriate when cooperative binding of many
chemoattractant molecules is required to trigger cellular
migration. The response $\gamma$ itself may also be
time-dependent. Also, note that not only might the production of
chemoattractant $b$ encounter a delay, the cell response embodied by
$\gamma$ might be also delayed.  Although such delayed and nonlinear
responses may result in intrinsically rich signaling behavior, for the
sake of simplicity, and in order to analyse our results with as few
free parameters as possible, we assume the ``force'' on the cell is
proportional to the local chemoattractant gradient {\it i.e.},

\begin{equation}
\dot{\R}(t) = \gamma \nabla n_{b}(\r,t)\big|_{\r = \R(t)}, 
\label{RDOT1}
\end{equation}

Stochastic effects due to low $n_{a}, n_{b}$ concentrations or other
random effects within the signaling process \cite{WINGREEN} can also
be easily incorporated component by either considering a randomly
varying $\gamma(t)$ and/or by adding a random noise to Eq. \ref{RDOT}
or \ref{RDOT1}. The additive noise is equivalent to endowing the cell
with Brownian motion. Nonetheless, we show that even in the ideal case
of perfect signaling and constant $\gamma$, novel target selection
phenomena arise.

In the following analysis, we define dimensionless parameters by
measuring length in units of the initial separation $R_{*}$ between
the probe and its farthest target, and time in units of
$(R_{*})^{2}/D_{a}$.  In spatial dimension $d$, the dimensionless
equations are identical in form to Eqs. \ref{DIFFA}-\ref{RDOT} except
with $D_{a}\equiv1$, and the terms $D_{b}, \mu_{a,b}, k_{j}, \gamma$,
and $F(t)$ replaced by $D_{b}/D_{a}$, $\mu_{a,b}R_{*}^{2}/D_{a}$,
$k_{j}/(D_{a}R_{*}^{d-2})$, $\gamma/(D_{a}R_{*}^{d})$, and
$R_{*}^{2}F(t)/D_{a}$, respectively.

The solutions to Eqs.~\ref{DIFFA} and \ref{DIFFB} can be solved in
terms of Green's functions and the cell position $\R(t)$. Upon
substitution of $n_{b}$ into Eq. \ref{RDOT1}, we find a self-consistent
nonlinear equation for the cell position

%\begin{widetext}
\begin{equation}
\fl \dot{\R}(t) = \gamma \sum_{j}k_{j} \int_{0}^{t}\dd t' \int_{0}^{t'}\dd t''
F(t'') G_{a}(\R(t'')-\R_{j};t'-t''){\bf H}_{b}(\R(t)-\R_{j};t-t'),
\label{RDOT2}
\end{equation}
%\end{widetext}
%
where $G_{a}(\r;t) = (4\pi
t)^{-d/2}e^{-r^2/4t}e^{-\mu_{a}t}$, and 

\begin{equation}
\begin{array}{l}
{\bf H}_{b}(\R(t)-\R_{j};t-t') \equiv \nabla
G_{b}(\r-\R_{j};t-t')\big|_{\r=\R(t)} \\[13pt] 
\displaystyle \hspace{1cm}= \displaystyle {2\pi(\R_{j}-\R(t))e^{-\vert
\R(t)-\R_{j}\vert^2/[4D(t-t')]}e^{-\mu_{b}(t-t')} \over [4\pi D
(t-t')]^{d/2+1}}.
\end{array}
\end{equation}
Although the model equations are linear in $n_{a}, n_{b}$, the moving
source of probe chemical renders the problem intrinsically nonlinear
and not amenable to analytic treatment.  Since bounds and analytic
expressions for $n_{b}(\r,t)$ and $\R(t)$ can be found only in special
cases (such as $F(t)\propto \delta(t)$), we will solve for the cell
trajectory by either numerically integrating Eq. \ref{RDOT2}, or by
directly numerically solving Eqs.~\ref{DIFFA}-\ref{RDOT} on a fixed
lattice using a stable backward-time, central space scheme with step
sizes $\Delta x = \Delta t = 10^{-3}$. in this case, the system
boundary is chosen to be far enough away from the targets as to be
irrelevant.  We have verified that our results do not depend on the
numerical approach employed.

\section{Results and Discussion}

We shall study our model predominately by solving either
Eqs. \ref{DIFFA}-\ref{RDOT} or Eq. \ref{RDOT2} numerically, and
exploring the qualitative features of the chemolocation
mechanism. However, in certain physical limits, we find analytical
relationships useful in describing target selection.

\subsection{Single Target}

In biological media or in laboratory realizations \cite{CLUZEL},
diffusion often occurs in confined or ramified geometries.  For
chemotaxis across capillaries, or across percolating paths, the
effective dimensionality $d$ of the diffusion process may be smaller
than the spatial dimension.  For simplicity, we first explore the
qualitative behavior of the one-dimensional ($d=1$ and ${\bf R}(t)
\equiv X(t)$) version of our model with initial condition
$n_{a}(x,0)=n_{b}(x,0)=X(0)=0$, and $\gamma=1$. As a demonstration of
the chemolocation mechanism, consider a cell moving towards a single
target under different probe release protocols $F(t)$. Different
strategies of probe release qualitatively influence the ability of the
cell to reach the target.

%
%%%%%%%%%%%%%%%%%%%%%%%%%%%%%%%%%%%%%%%%%%%%%%%%%
\begin{figure}[htb]
\begin{center}
\includegraphics[width=4.5in]{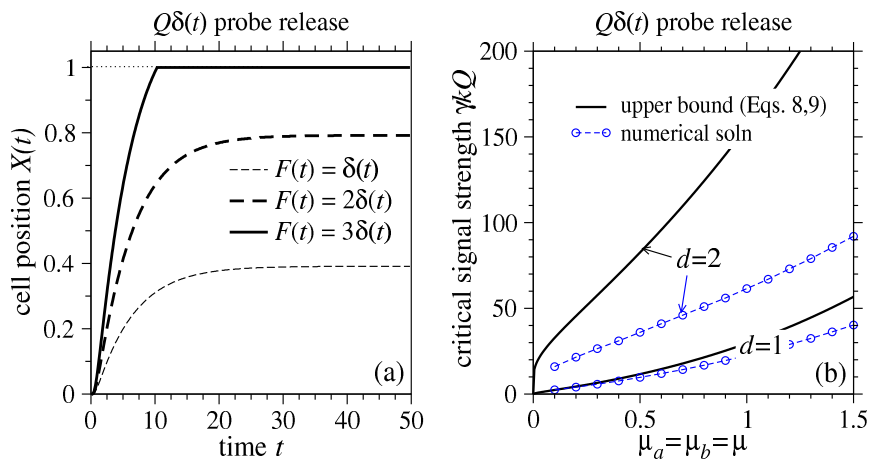}
\end{center}
\caption{\baselineskip=12pt Chemolocation to a single target using a
single $\delta$-function probe release. (a) Motion of the probe
towards a single target in $d=1$ using three different
$\delta$-function release intensities.  Delta-functions were
approximated with releases of duration $\dd t = 0.1$.  Parameters used
were $D=\gamma=k=1$ and $\mu_{a}=\mu_{b}=0.1$.  (b) Critical signaling
strength $\gamma k Q$ as a function of chemical decay rates $\mu =
\mu_{a}=\mu_{b}$.  Results from numerical solutions and the upper
bound Eq. \ref{UPPERBOUND} are shown for both $d=1$ and $d=2$.}
\label{FIG2}
\end{figure}
%%%%%%%%%%%%%%%%%%%%%%%%%%%%%%%%%%%%%%%%%%%%%%%%%
%

Figure~\ref{FIG2} shows the trajectories and maximum distance traveled
by the cell when a single $\delta$-function probe is emitted at $t=0$.
Here, and in the rest of the paper, the ``$\delta$-function'' is
approximated by a narrow square pulse release of duration $\dd t =
0.1$ and intensity $Q$.  The cell starts to move only after a short
delay during which some of the probe has reached the target, and the
converted chemoattractant has diffused back to the cell. For a single
impulse release of probe $a$, the velocity of the cell towards the
target is initially high but eventually goes to zero since the system
runs out of the chemoattractant $b$ once the single pulse of probe $a$
has dissipated. Thus, for a modest ({\it e.g.}  $Q=1,2$) single-pulse
release in Fig.~\ref{FIG2}(a), the cell moves only part of the
distance to the target. In the low mobility (small $\gamma$) limit, an
approximation to the total travel distance $X_{\infty}$ after probe
release in a single pulse can be obtained by setting ${\bf
R}(t'')={\bf R}(t)=0$ in Eq. \ref{RDOT2}.  In one-dimension ($d=1$),
integrating the left-and-side of Eq. \ref{RDOT2} yields
$\int_{0}^{\infty} \dot{X}(t)\dd t = X_{\infty}$. Explicit evaluation
of the time integral of the right-hand-side of Eq. \ref{RDOT2} shows
that

\begin{equation}
X_{\infty} >  {Q\gamma \over 4D\sqrt{\mu_{a}}}\sum_{j} k_{j}{X_{j} 
\over \vert X_{j}\vert}
e^{-\left(\sqrt{\mu_{a}}+\sqrt{\mu_{b}/D} \right)
\vert X_j \vert}.
\label{APPROX}
\end{equation}
When there is only a single target at $X_{1} = 1$,  Eq. \ref{APPROX}
gives a lower bound for $X_{\infty}$, implying that

\begin{equation}
\gamma k Q 
\geq 4D\sqrt{\mu_{a}}e^{\sqrt{\mu_{a}}+\sqrt{\mu_{b}/D}} \quad (d=1)
\label{UPPERBOUND}
\end{equation}
is a sufficient condition for the cell to reach the target.  The
analogous sufficient conditions for insuring the cell reaches a single
target in $d=2$ and $d=3$, are 

\begin{equation}
\gamma k Q \geq
{4\pi^{2}\over \sqrt{\mu}K_{0}(\sqrt{\mu})K_{1}(\sqrt{\mu})} \quad (d=2),
\label{UPPERBOUND2}
\end{equation}

\noindent and 

\begin{equation}
\gamma k Q \geq {16\pi^{2}e^{2\sqrt{\mu}}\over \sqrt{\mu}+1} \quad (d=3),
\label{UPPERBOUND3}
\end{equation}
respectively.  These ``critical'' values of the signalling strength
$\gamma k Q$ represent the analytically obtainable lowest values above
which the target is acquired with certainty.  For $d=2,3$, these
conditions derived from the ${\bf R}(t'')={\bf R}(t) = 0$
approximation are independent of the relative diffusivity $D$. In
higher dimensions, the spreading of probe and chemoattractant
concentrations in more severe so that under the approximation ${\bf
R}(t'')={\bf R}(t) \approx 0$, the stationary cell experiences a
significantly diminished gradient compared to that which it
experiences if it had moved.  Chemical decay $\mu$ also diminishes the
signal.  Therefore, the signalling strength required for reaching the
target is significantly increased in this ``adiabatic'' approximation,
especially for large $d$ or $\mu$. Fig.~\ref{FIG2}(b) plots the
critical values (for $d=1$ and $d=2$, $d=3$ not shown) of the
signaling strength $\gamma k Q$, from both simulations and
Eq. \ref{UPPERBOUND}, as functions of $\mu = \mu_{a} = \mu_{b}$.
Indeed, the sufficient condition Eq. \ref{UPPERBOUND} for $d=1$
provides the tightest bound.

\begin{figure}[htb]
\begin{center}
\includegraphics[width = 4.5in]{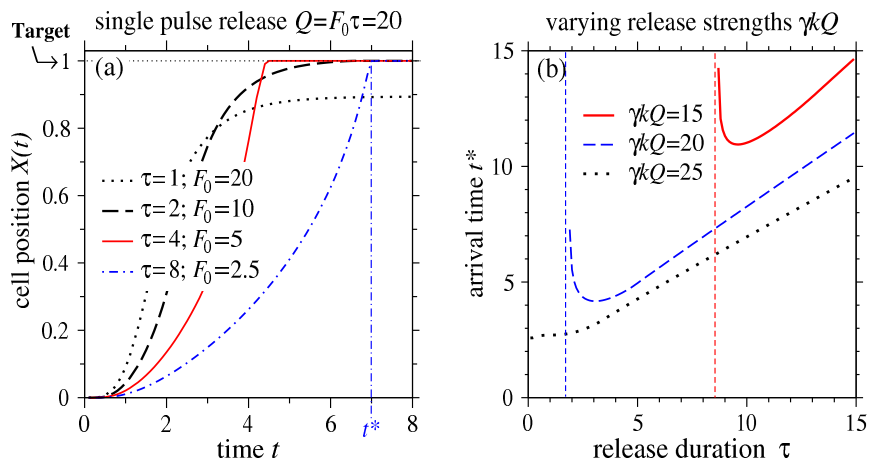}
\end{center}
\caption{\baselineskip=12pt Chemolocation to a single target using a
probe pulse of duration $\tau$.  (a) Trajectories of a cell to a
single target when a fixed, total amount of probe $Q$ is released at a
constant rate, $F_0$ for a finite time, $\tau$.  (b) Arrival times
$t^*$ of a cell to a single target are plotted as a function of pulse
duration $\tau$.  Within each curve, the total amount of probe
released, $Q$ was held constant.  In both plots, parameters used were
$D = \gamma = k = \mu_A=\mu_B=1$.}
\label{FIG3}
\end{figure}

The likelihood of arrival to a single target can be enhanced not only
by increasing the signaling strength $\gamma k Q$, but also by
releasing multiple pulses (as will be shown in Fig. \ref{FIG4}(b)) and
by releasing probe more slowly.  Consider the case where the cell
contains a fixed amount of probe chemical $a$. How should the cell
release this fixed amount of probe to best reach the target?  Suppose
the release occurs in a single pulse of duration $\tau$ and amplitude
$F_{0}=F(t\leq \tau)$ such that the total amount $Q=F_{0}\tau$ of
chemical released is constant.
%XXXXXXXXXXX REWRITE OF FIG 3 XXXXXXXXXXXXXXXXXX
Fig.~\ref{FIG3}(a) shows trajectories for a cell that releases a fixed
amount, $Q$, of chemical $a$ at a constant rate $F_0$ over different
lengths of time $\tau = Q/F_0$.  For $F_{0}=20$ and $\tau = 1$, the
large magnitude, short duration release allows the cell to travel only
approximately 90\% of the way to the target. In such cases, the cell's
velocity may reach a high value; however, the cell does not reach the
target before all chemical signals have dissipated. For lower
intensity, but longer duration releases, the cell is better able to
reach the target.  Qualitatively, this can be understood by noting
that a single $\delta$-function release gives a lower bound on the
distance $X_{\infty}$ traveled. If the probe is released more slowly,
the cell moves slightly closer to the target, amplifying the effect of
the probe chemical that is released at later times, because this
portion can reach the cell more quickly reducing its decay.

In the limit $\tau \rightarrow \infty$, with fixed $Q = F_{0}\tau$, we
can find an {\it upper} bound $X_{\infty}^{\rm max}$ for the distance
travelled to a single target by integrating Eq. \ref{APPROX} and
summing the total distances travelled for each independent increment
$\dd Q = F_{0}\dd\tau$ of probe release. Assuming $\mu_{a} = \mu_{b} =
\mu$, $D=1$, and $X_{1} = 1$ for simplicity, we find for $d=1$,

\begin{equation}
X_{\infty}^{\rm max} = 1- {1\over
2\sqrt{\mu}}\ln\left(e^{2\sqrt{\mu}}-{\gamma k Q \over 2 D}\right),
\end{equation}
valid for $0\leq \gamma k Q \leq 2D(e^{2\sqrt{\mu}}-1)$. Therefore,
even for infinitely slow release of a fixed amount $Q$ of probe
chemical, a necessary condition $\gamma k Q \geq 2D(e^{2\sqrt{\mu}}-1)$
remains, even if the cell can detect arbitrarily low concentration
gradients of chemoattractant.

Even if the cell has sufficient probe to reach the target, if the
probe is released very slowly, the time required for the cell to reach
the target may be large. Fig. \ref{FIG3}(b) plots the time $t^{*}$ it
takes for the cell to reach the target, as a function of the pulse
duration $\tau$.  For large signaling strength $\gamma k Q$, the cell
reaches its target for all $\tau$, and reaches its target most quickly
for very short pulse durations, $\tau$.  When the signaling strength
$\gamma k Q$ is decreased, the cell reaches its target only when
$\tau$ is greater than some critical value, denoted by the vertical
asymptotes (dashed lines) in Fig. \ref{FIG3}(b).  Additionally, there
is a value of $\tau$ which minimizes the arrival time $t^{*}$, and
approximately satisfies the condition $\tau \approx t^*$, indicating
that the cell reaches its target in the shortest time when probe
chemical $a$ is released over a period $\tau$ nearly equal to its
travel time.  If we think of probe $a$ as an effective ``fuel'' that
drives the motion of the cell, it is not ideal to have $\tau \ll t^*$,
because if chemical $a$ stops being released well before the cell
reaches its target, the cell will have to rely on a residual, decaying
concentration field $b$ to reach its target.  On the other hand, if
$\tau \gg t^{*}$, the cell will continue to release chemical $a$ after
reaching its target, when a faster release of $a$ would have allowed
the cell to reach its target more quickly.

%%%%%%%%%%%%%%%%%%%%%%%%%%%%%%%%%%%%%%%%%%%%%%%%%
%%%%%%%%%%%%%%%%%%%%%%%%%%%%%%%%%%%%%%%%%%%%%%%%%

\subsection{Multiple Targets}

\begin{figure}[htb]
\begin{center}
\includegraphics[width = 6in]{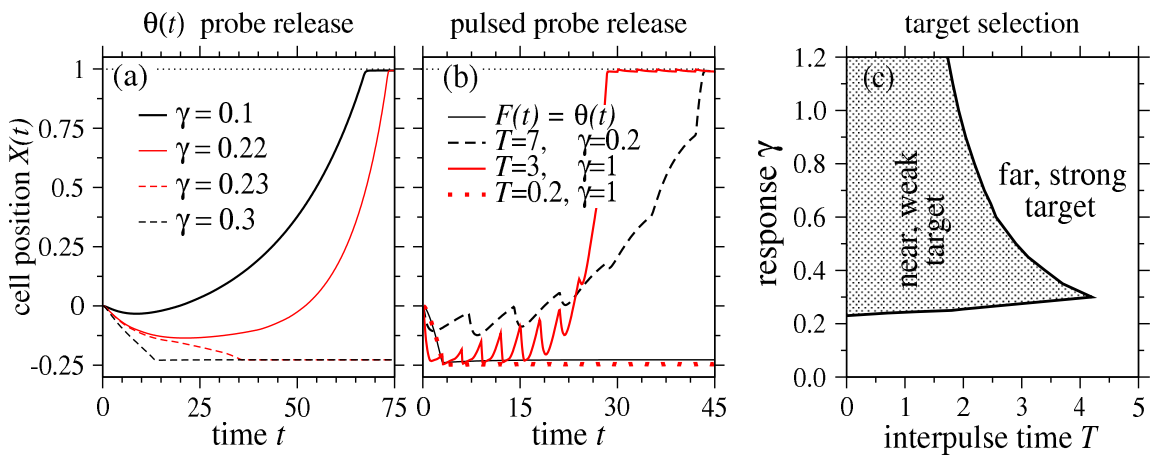}
\end{center}
\caption{\baselineskip=12pt Target selection as a function of
chemotactic response $\gamma$ and probe release frequency $1/T$. (a)
Response-dependent target selection showing a transition from
selecting the further, stronger, target to selecting the weaker,
nearer target for a Heaviside release protocol $F(t) = \theta(t)$ and
a varying response $\gamma$. When $\gamma \gtrsim 0.235$, the near
target is selected, while the far target is selected for smaller
$\gamma$. (b) Frequency-dependent target selection, with release
protocol $F(t) = T \sum_{i} \delta(t - iT)$ and varied response
strength $\gamma$.
%where $\alpha$ is the time-averaged release rate.  
The weaker nearer target is selected for constant chemical release
$F(t) = \theta(t)$ and very high frequency pulsed release $F(t) = 0.2
\sum_{i} \delta(t - 0.2 i)$, while the further stronger target is
selected for pulsed release at intermediate frequencies (corresponding
to interpulse intervals $T=3, 7$). These trajectories may first arrive
at the nearer target depending upon probe chemical release
amplitude. (c) Under the release function $F(t) = T\sum_{i}
\delta(t-iT)$, regions of $\gamma-T$ space where the cell reaches the
nearer (and weaker), or farther (and stronger) target are shown.  When
$T \gtrsim 4$, the cell always reaches the far target.  For smaller
values of $T$, the cell reaches the far target for small and large
values of $\gamma$, and reaches the near target for intermediate
values of $\gamma$.  For all plots, the target strengths are $k_1 =
0.5$, $k_2 = 1$, the decay rates are $\mu_a = \mu_b = 0.1$ and $D =
1$. The initial separations between the cell and the targets are
$\vert X_1\vert = 0.25$ and $\vert X_2\vert =1$ respectively.}
\label{FIG4}
\end{figure}

We now illustrate the mechanism of frequency and response
strength-dependent target selection among multiple targets.  First
consider the one-dimensional case with a cell and two targets. The
closer target (at $X_{1}=-0.25$) is assigned a
probe-to-chemoattractant production rate $k_{1}=0.5$, while the
farther target at $X_{2}=1$ has a larger production rate $k_{2}=
1$. Fig.~\ref{FIG4}(a) shows that when the release of probe chemical
is in the form of a Heaviside function, target selection can be
controlled by the cell's response, $\gamma$. For the parameters
chosen, if $\gamma = 0.1$, the far target is chosen. As the strength
of the response is increased beyond approximately $0.225$, the near
target is selected by our chemotactic mechanism. When the cell's
response, $\gamma$ is large, the cell is initially pulled strongly
towards the near target.  Because the distance to the far target
increases substantially before the signal from this target reaches the
cell, the signal strength diminishes and is insufficient to pull the
cell back towards the far target.  When the cell's response is small,
the cell is unable to move much towards the near target before the
stronger signal from the far target reaches the cell, and the cell
ultimately gets pulled to the far target.

%%%%%%%%%%%%%%%%%%%%%%%%%%%%%%%%%%%%%%%%%%%%%%%%%
%%%%%%%%%%%%%%%%%%%%%%%%%%%%%%%%%%%%%%%%%%%%%%%%%

Target selection can depend not only on probe release intensity and
response strength, but is also modulated release
frequency. Fig.~\ref{FIG4}(b) illustrates target selection when the
probe chemical is released either as a Heaviside function $\theta(t)$,
or as a series of pulses $F(t) = T \sum_{i}\delta(t-iT)$ with an
interpulse interval $T$ and varying response $\gamma$. This form of
$F(t)$ compares release protocols with equal long-time-averaged probe
release. For the parameters chosen, the constant release $F(t) =
\theta(t)$ results in the cell arriving at the near target.  For
pulsed release with interpulse interval $T=3,7$, the cell initially
moves towards, and depending on the pulse intensity, may first reach
the near target before eventually being pulled to the farther,
stronger target. If the release frequency is increased even further,
(interpulse interval $T=0.2$, red dotted curve), the trajectories will
again arrive at the closer, weaker target. Since the underlying
processes are dissipative, at very high frequencies, the cell cannot
respond fast enough to distinguish the pulses and a rapid succession
of pulses is equivalent to an effective-amplitude, constant emission.

In Fig.~\ref{FIG4}(c), we show a phase diagram indicating the regions
of $\gamma$-$T$ space in which we expect the cell to go to the near,
weak target, or to the far, strong target.  When the interpulse time
$T$ is small and the cell's response $\gamma$ is small, it will go to
the far target.  For $\gamma > 0.225$, the near target is selected.
These results are consistent with those shown in Fig.~\ref{FIG4}(a).
When the interpulse interval $T$ is large, the cell will reach the far
target for any response strength $\gamma$.  In this example, the cell
will reach the far target for any value of $T\gtrsim 4$.  When $T$
falls in the interval $0<T<4$, the cell will reach the far target when
$\gamma$ is either small or large, but will reach the near target at
intermediate values of $\gamma$.  When the cell reaches the far target
and $\gamma$ is large, initial chemoattractant pulses quickly pull the
cell towards the near target.  For several pulses, alternating
chemoattractant waves from the far and near targets will pull the cell
away from, then back towards, the near target.  The cell appears to
``bounce'' around the near target.  Eventually, a wave of
chemoattractant from the far target will dislodge the cell from the
near target, and the next wave of chemoattractant from the near target
will be insufficient to return the cell to the near target.  After
this, each subsequent pulse brings the cell closer to the far target,
until the cell reaches this target.  The trajectory with parameters
$\gamma = 1$, $T = 3$ illustrates this qualitative behavior in
Fig.~\ref{FIG4}(b).  When the cell with small $\gamma$ reaches the far
target, it does so without first reaching the near target.  The
trajectory with parameters $\gamma = 0.2$, $T = 7$ in
Fig.~\ref{FIG4}(b) illustrates this qualitative behavior.

A more quantitative understanding of the target selection phenomenon
can be found in the small $\gamma$ limit, where the cell does not move
much under the influence of a single probe pulse (such as the
trajectory corresponding to $\gamma = 0.2$ and $F(t) =
7\sum_{i}\delta(t-7i)$ in Fig.~\ref{FIG4}(b)).  If the cell does not
move appreciably under the influence of a single probe pulse, we can
use Eq. \ref{APPROX} to approximate the asymptotic distance
traveled by the cell as a result of one single probe pulse released at
$t=0$. If $X_{\infty} < 0$, subsequent probe pulses will be released
when the cell is closer to the left target, and the cell will
eventually incrementally move leftward. If $X_{\infty} >0$, the cell
will ultimately arrive at the right target. Therefore, $X_{\infty}=0$
defines an approximate boundary for selection between two targets in
$d=1$.

\begin{figure}[h]
\begin{center}
\includegraphics[width=3.25in]{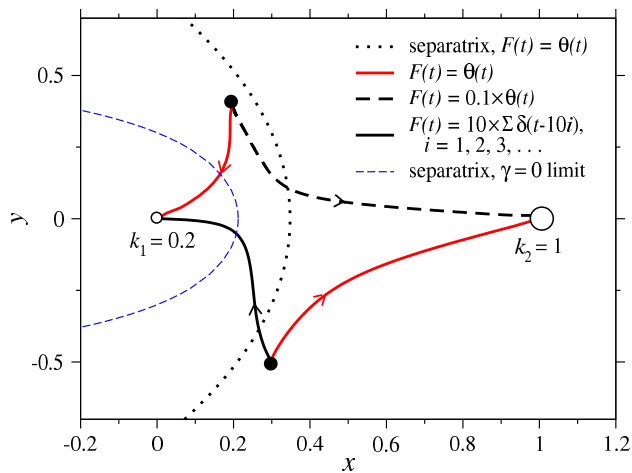}
\end{center}
\caption{\baselineskip=12pt Target selection in $d=2$ depends on probe
release protocol. The separatrix corresponding to $\theta(t)$ probe
release (dotted line) divides the space such that trajectories
(solid red lines) originating from points to the right of the dotted line
arrive at the stronger target (big open circle) while those starting
from the left arrive at the weaker target (small open circle).
% under constant
%probe chemical release protocol $F(t) = \theta(t)$. 
Amplitude-dependent target selection is shown by using $F(t) = 0.1
\times \theta(t)$ (black dashed line), leading to a different target
than when $F(t) = \theta(t)$.  Likewise, frequency-dependent target
selection is demonstrated using $F(t) = 10 \sum_{i} \delta(t - 10i)$
(black solid line with a unit average probe release) where the cell
selects the nearer target when started in a region that would normally
lead to selection of the stronger, farther target when $F(t) =
\theta(t)$. For comparison, the separatrix for continually constant
probe release $F$ computed by integrating along the ridge of the
static field $n_{b}(\r)$ is shown by the thin blue curve. Parameters
used were $D = k_1 = 1$, $k_{2}=0.2$, $\mu_a = \mu_b = 0.001$ and
$\gamma = 10$. The numerical solution to Eqs. \ref{DIFFA}-\ref{RDOT}
in $d=2$ were found using the Peaceman-Rachford algorithm.}
\label{Separatrix-Target-Selection-Fig}
\end{figure}
%%%%%%%%%%%%%%%%%%%%%

Generalizing the phenomenon to $d$ dimensions, we expect that for each
release sequence $F(t)$, there will be at least one $d-1$ dimensional
surface that separates trajectories that evolve to different targets
starting from a given initial position on the $d$ dimensional
manifold. Figure~\ref{Separatrix-Target-Selection-Fig} shows
trajectories of cells searching and selecting between two targets in
$d=2$. The separatrix associated with the Heaviside probe release
$F(t) = \theta(t)$ is indicated by the dotted curve that divides the
space such that trajectories originating from the right/left of this
line are led to the stronger/weaker target respectively.  For
comparison, the separatrix for constant probe release for all times
(corresponding to the $\gamma \rightarrow 0$ limit) computed by
integrating along the ridge of the static field $n_{b}(\r)$ is shown
by the thin blue curve, highlighting its sensitivity to $F(t)$. We
then start the cell within regions where one target is clearly
selected when the probe chemical is released with $F(t) = \theta(t)$,
but with either a diminished released rate or with pulsed release,
such that the average rate of chemical released per unit time is 1.
As in the $d=1$ case we find that for Heaviside function release, a
larger release rate favors the weaker, closer target, while a smaller
release rate favors the stronger, farther target (thick dashed curve).
In one dimension we find that slow pulsed release favors the far,
strong target over the near, weak target.  In contrast, for $d=2$, we
find that slow pulsed release favors the near target (solid black
curve).  This may be understood as follows. In $d=1$, pulsed release
causes the cell to quickly go towards the near target, but a
subsequent wave of chemoattractant from the far target caused the cell
to ultimately select the far target.  In two dimensions, the radial
divergence of the concentration fields renders the signal from the far
target insufficient to pull the cell from the weaker target. In higher
dimensions, the radial spread is stronger, and the frequency and
$\gamma$-dependent selection mechanism is significantly mitigated.

While nontrivial target selection phenomena still arises in $d=2$,
many cellular chemotactic responses, such as development, occur in
$d=3$ where the concentration fields spread significantly and
attenuate signalling responses. However, in many situations, the
medium in which the concentration fields $n_{a}$ and $n_{b}$ diffuse
is heterogeneous.  A common biological example is the extracellular
cellular matrix and the existence of intervening cells and tissues.
Diffusive transport in a random medium can also be described by
diffusion in a lower, effective dimension. Furthermore, media that is
ramified can transport signals in a one-dimensional manner along the
main percolating path once side branches have saturated with probe
and/or chemoattractant. Therefore, signatures of our proposed mechanism
may nonetheless arise in $d=3$.

Although our results are based on a simplified model of response
(Eq. \ref{RDOT1}) where $U[n_{b}(\r,t)] \propto n_{b}(\r,t)$, and
neglect noise of memory effects in $\gamma$, they provide clues into
how more complex models might behave. For example, in the limit of a
motility response that requires highly cooperative binding of multiple
chemoattractants (large Hill coefficient $\alpha$), the motion of the
cell would be appreciable only when a very narrow range of gradients
in $n_{b}$ cross the cell. High cooperativity imposes a switching
response which would lead to saltatory movements of the cell. While
larger velocities can be attained (because $\nabla U[n_{b}]$ can be
larger is $\alpha$ is large) the time scale over which $\nabla
U[n_{b}]$ is large is short.  We expect the net effect to be
quantitative and that most our qualitative results will persist.

Delayed response, or directional memory, may arise from receptor
adaptation and other intermediate steps in the cell signaling pathway
\cite{TU}, and may give rise to behavior that is more difficult to
intuit without numerical computations. However, simple conclusions in
certain limits can be motivated by comparing the delay time, or
directional correlation time with the arrival times of the gradients
$\nabla U[n_{b}]$ from the targets.  For example, if directional
memory is longer than the difference in times it takes for the $n_{b}$
signal to propagate from the close and far targets, the cell would be
less sensitive to the signal from the far target. This would have the
effect of biasing selection of the near target, provided the release
frequency is sufficiently high.

One example of a biologically-motivated variant of our model worth
further discussion is ``autocrine'' signalling. Autocrine signalling
is often relevant to bacterial aggregation, where the probe is
identical to the chemoattractant ($a=b$). Within our mathematical
framework, the autocrine mechanism is described by replacing
$n_{b}(\r,t)$ with $n_{a}(\r,t)$ in Eqs. \ref{DIFFB} and \ref{RDOT1},
and setting $\mu_{b}=\mu_{a}$ and $D_{b}=D_{a}$. This model is also
novel in that its naive continuum limit does not map onto existing PDE
descriptions such as Keller-Segel type models.  In the Keller-Segel
model, the production of autocrine chemoattractant is simply
proportional to the local cell density and is independent of the
probing signal the cells receive. Our qualitative results for
paracrine signalling also hold in the autocrine signalling problem,
which can be understood by considering the $n_{a}$ profile generated
by the cell and that generated by the target. In the autocrine case,
when the cell responds to $n_{a}$, its self-produced $n_{a}$ field is
maximal near the point where it is released.  Therefore, the
self-produced $n_{a}$ tends to keep the cell from moving. When the
gradient in $n_{a}$ generated from the target is finally felt, the
cell will move towards the target; however, typically by this time,
the self-generated $n_{a}$ has diffused and/or dissipated away such
that it's ``restoring force'' is weak. Therefore, we expect that all
else being equal, the replacement of $n_{b}$ by $n_{a}$, leads to an
overall slightly suppressed chemotactic response.

Finally, analysis of the effects of stochastic responses, implemented
through additive Brownian terms to Eq. \ref{RDOT} or \ref{RDOT1}, or
through a time random response coefficient $\gamma(t)$, is beyond the
scope of this study. However, it has been shown that in a
two-dimensional autocrine mechanism, stochastic effects and chemical
decay keep the cells diffusive~\cite{Grima:05}, despite the tendency
for all cells and targets to aggregate.  Therefore, one might expect
that target selection to be suppressed in the presence of sufficient
noise.

%A quasi $2$-d system in which a cell crawls along a line to
%a $2$-d plane {\it e.g.} as in tissues would also give pronounced
%target selection as a function of probe frequency.
%%%%%%%%%%%%%%%%%%%%%%%%%%%%%%%%%%%%%%%%%%%%%%%%%
%%%%%%%%%%%%%%%%%%%%%%%%%%%%
\section{Summary \& Conclusions}

In conclusion, we have proposed a model for dynamic, multistep
chemotaxis that involves chemical communication between cell and
targets. Our analysis shows that signalling agents can select among
potential targets by controlling the amount of, and frequency at which
probe chemical is released. Since our moving source problem is
intrinsically nonlinear, we employed numerical calculations to provide
evidence of a critical target-switching release amplitude, as well as
a window of response strengths $\gamma$ within which a cell chooses
the nearer, weaker target over the farther, stronger target that is
selected outside this window.  This effect arises from a nonlinear
interplay between diffusion, decay, and chemoattractant production and
depends on the frequency of release.  Moreover, we found that for the
conditions studied, a constant probe release leads to acquisition of
the weaker, nearer target, while low frequency spike release leads to
acquisition of the stronger, farther target.  At higher frequencies,
the cell again approaches the weaker, nearer target since high
frequency and constant release give rise to similar spatial-temporal
probe distributions due to the averaging nature of the underlying
diffusive physics.  Our numerical experiments have shown that target
selection is observed over a wide range of system parameters,
suggesting the chemolocation mechanism may be common in Nature.

Numerous variants of our basic model, such as mobile targets that
sense probe chemical, stochastic effects, and delays in the signaling
processes can be straightforwardly investigated. In the present work,
we assumed the simplest deterministic, instantaneous response in order
to avoid introducing excessive number of parameters.  Nonetheless, we
have argued that many extensions of our underlying model, except
sufficiently large noise, are likely to qualitatively preserve the
predicted results.

%\section{Acknowledgements}

\vspace{5mm}

The authors thank M. R. D'Orsogna for helpful discussions.  This
research was supported by grants from the NSF (DMS-0349195) and the
NIH (K25AI058672). SAN acknowledges support from an NSF Graduate
Research Fellowship.

%\end{article}

%%%%%%%%%%%%%%%%%%%%%%%%%%%%%%%%%
\end{document}